\documentclass[iop]{emulateapj}
\usepackage{graphicx}
\usepackage{hyperref}
\usepackage{natbib}
\usepackage{color}
\usepackage{mathtools}

\bibpunct{(}{)}{;}{a}{}{,}

\shorttitle{NLR-BLR in AGNs}
\shortauthors{Adhikari et al.}
\slugcomment{Accepted for publication in ApJ}

\begin{document}


\title{The intermediate line region in active galactic nuclei}


\author{T. P. Adhikari\altaffilmark{1}, A. R\'o\.za\'nska\altaffilmark{1}, B. Czerny\altaffilmark{2},
K. Hryniewicz\altaffilmark{1} and G. J. Ferland\altaffilmark{3,1}}
\email{tek@camk.edu.pl}
\altaffiltext{1}{Nicolaus Copernicus Astronomical Center, Polish Academy of Sciences, Bartycka 18, 00-716, Warsaw}

\altaffiltext{2}{Center for Theoretical Physics, Polish Academy of Sciences, Aleja
Lotnikow 32/46, Warsaw, Poland} 

\altaffiltext{3}{Department of Physics and Astronomy, The University of Kentucky, Lexington, KY 40506, USA} 

\begin{abstract}

We show that the recently observed suppression of the gap between the 
broad line region (BLR) and the narrow line region (NLR) in some AGN 
can be fully explained by an increase of the gas density 
in the emitting region. 
Our model predicts the formation of the intermediate line region (ILR) that is 
observed in some Seyfert galaxies by the detection of 
emission lines with intermediate velocity
full width half maximum (FWHM) $\sim$ 700 - 1200 km s$^{-1}$.
These lines are believed to be originating 
from an ILR located somewhere between the BLR and NLR. 
As it was previously proved, 
the apparent gap is assumed to be caused 
by the presence of dust beyond the sublimation radius.
Our computations with the use of {\sc cloudy} photoionization code, 
show that the differences in the shape of spectral energy distribution (SED) 
from the central region of AGN, 
do not diminish the apparent gap in the line emission in those objects.
A strong discontinuity in the line emission vs radius
exists for all lines at the dust sublimation radius. 
However,  increasing the gas density to $\sim$ 10$^{11.5}$ cm$^{-3}$ at the sublimation
radius provides the continuous line emission vs radius
and fully explains the recently observed lack
of apparent gap in some AGN. We show that such a high density is consistent 
with the density of upper layers of an accretion disk atmosphere.
Therefore, the upper layers of the disk atmosphere can 
give rise to the formation of observed emission line clouds. 
\end{abstract}



\section{Introduction}
Observations of active galactic nuclei (AGN) clearly show 
 different types of emission lines in their
spectra. Depending on the width of the emitted lines, 
they are classified as  narrow lines, 
with full width half maximum (FWHM) $\sim$ 500 km s$^{-1}$,
or broad lines, with 
FWHM $\gtrsim$ 2000 km s$^{-1}$.
This implies two different regions of line formation, i.e, narrow 
line region (NLR) and the broad line region (BLR)
respectively.
Although the physical 
conditions of these regions are quite well
understood
\citep{davidson1972,krolik1981,netzer1990,pier1995,dopita2002,groves2004a,
groves2004b,stern2014,baskin2014,czerny2011,czerny2015,stern2016},
there are still several issues to be explored. One of those problem
is the significant suppression of
the line emission in a region between the BLR and NLR in most AGN
\citep{osterbrock2006, boroson1992}. A leading explanation for this feature 
is the dust suppression introduced by \citet[hereafter NL93]{netzer1993}.
On the other hand,  recent observations of several objects 
indicate the existence of
an intermediate line emission region (ILR), between the BLR 
and NLR, which produces emission lines
with velocity FWHM $\sim$ 700 - 1200 km s$^{-1}$.
Using the statistical investigations of broad UV-lines in QSOs,
\citet{brotherton1994} discussed the ILR as an inner extension of the NLR.  
\citet{mason1996} found  evidence for an ILR with velocity
FWHM $\sim$ 1000 km s$^{-1}$ which produces a
significant amount of both permitted and
forbidden line fluxes in NLSy1 RE~J1034+396.  For the Sy1
NGC 4151, \citet{crenshaw2007} identified a 
line emission component with  width FWHM $=$ 1170 km s$^{-1}$, most probably
originating between the BLR and NLR. Detailed spectral analysis of large number of 
{\it SDSS} sources have revealed the presence of intermediate component of 
line emission with velocity width in between that of broad and narrow components \citep{hu2008a,hu2008b,zhu2009}.
Additionally, \citet{crenshaw2009}
detected  ILR emission with width of 680 km s$^{-1}$ in the historically 
low state spectrum of NGC 5548 obtained with
the Space Telescope Imaging Spectrograph on the 
{\it Hubble Space Telescope}. Using the {\it HST/ FOS} spectra of quasar OI 287,  \citet{li2015} reported
 the detection of intermediate width emission lines originating at 
 a distance of $\sim$ 2.9 pc from the central black hole.
The question remains unanswered, is there any 
separation between gas responsible for broad 
and narrow line emission in different types of 
AGN? What is the physical mechanism
leading to the existence of this gap in some source, and 
why the gap is not observed in all of them?

Photoionization calculations provide a powerful 
tool to model the line emission 
from gas clouds powered by the
continuum radiation from the nucleus. 
Detailed analysis  has  shown that BLR clouds 
are denser and located closer to the AGN center, 
while the NLR clouds have lower density and are more distant
\citep[see for details][]{osterbrock2006,netzer1990}.
Nevertheless, the lack of significant line emission from 
the intermediate zone between the BLR and NLR 
is not naturally explained by photoionization models. 

The presence of dust complicates the physics by introducing 
additional processes 
\citep[and references therein]{ferland1979,baldwin1991,vanhoof2004} 
in the radiation-matter interactions 
which has to be treated properly in  simulations of 
the gas emission.
Observations have shown the existences
of dust in the NLR for almost all types of AGN
\citep[]{derobertis1984,ferland1986,wills1993}.  
However, it has been argued that the BLR clouds are devoid of dust
since there is a lack of depletion of refractory elements in the gaseous phase \citep{gaskell1981}.
Usually, it is believed that, the radiation energy is so high 
in BLR, that the dust if present sublimes and no longer 
survives there \citep{czerny2011}.
Reverberation mapping studies show that the BLR clouds are located at a distance smaller by a factor 4 to 5 than the hot dust emission
\citep{suganuma2006,koshida2014}.
On the other hand, \citet{nenkova2008} has shown that the  closest region where dust can survive in the full radiation field is the face
of the dusty torus, located approximately at 0.4 pc for AGN of typical luminosity (see their Eq.~1).
While Eq.~1 of \citet{nenkova2008} gives the smallest radius
 at which the dust absorption coefficient reflects the full grain mixture, the largest
grains survive to closer radii, where they are presumably detected
by the reverberation measurements.

The observed apparent gap in the line emission region between BLR and NLR was explained
by the dust content in NLR clouds by NL93. 
The authors  calculated  emission from 
a continuous radial distribution of clouds 
extending from the BLR to the NLR, using the numerical photoionization code
{\sc ION} \citep{rees1989}.
Their assumption was that the dust is present in NLR 
but sublimes in a higher ionization region, 
around the BLR. The authors have shown that the reduced line
emission vs radius is a result of the dust extinction of
ionizing radiation as well as the dust destruction of 
the line photons. The dust absorption
becomes more efficient with decreasing  radial 
distance from the nucleus, and gives rise to an empty intermediate
region, where gas is present but the line emission is heavily suppressed. 
The dust fully sublimates at smaller radii, 
and  line emission increases 
dramatically, by about an order of magnitude, giving rise to the BLR.
NL93 successfully demonstrated the apparent gap in the line emission vs radius, 
although their result was obtained for a
particular spectral energy distribution (SED) typical for Sy1 AGN 
\citep{Mathews1987}, and for a specific value of the density 
$n_{\rm H} = 10^{9.4}$ cm$^{-3}$ at the sublimation radius at $\sim~0.1$ pc.

The aim of this paper is to explain the disappearance of an
apparent gap in line emission seen for some sources 
within the framework of NL93 model.
We use the photoionization code {\sc cloudy}, version 13.03
\citep[][]{ferland2013} to determine line luminosities 
for different shapes 
of the incident radiation representative for various sub-classes
of AGN, as measured from recent observations: 
Sy1.5 galaxy Mrk 509 \citep{Kaastra2011},
Sy1 galaxy NGC 5548 \citep{Mehdipour2015}, and 
NLSy1 galaxy PMN J0948+0022 \citep{ammando2015}. 
In addition, we also use a flexible
parametric shape of SED as a  
Band function \citep{band1993}.
All our results presented below focus on the five major emission 
lines usually detected in those objects: 
H${\beta}$ ${\lambda}$4861.36 \AA, He~II ${\lambda}$1640.00 \AA, 
Mg~II ${\lambda}$2798.0 \AA, C~III] ${\lambda}$1909.00 \AA 
~and [O~III] ${\lambda}$5006.84 \AA.

We show, that the various shapes of SED used in our computations
with the model parameters taken from NL93 do not remove apparent gap in the 
line emission vs radius. For lines considered by us, we always obtain 
apparent gap similar to the result of NL93.  
Therefore, the observed properties of ILR cannot be explained only by considering the
different shapes of radiation illuminating the gas and dust in our simulations.
However, increasing the gas density at the sublimation
radius yields to the continuous line 
emission vs radius and fully explains the observed emission from the ILR.

The structure of this paper is as follows:
Section~\ref{sec:parameters} describes the parameter 
set up of our photoionization simulations. 
In Section~\ref{sec:sed_dependence}, we
present the major line luminosity radial profiles for 
various SEDs. The same line emission vs radius, but 
for different gas densities, are presented 
in Section~\ref{sec:density_dependence}.
Finally, the discussion and conclusions of our work are described 
in Sections~\ref{sec:disc}~and~\ref{sec:con}.

\section{The model and its parameters}
\label{sec:parameters}

We consider the continuous distribution of optically thick spherical clouds
above an accretion disk, 
placed at different radial distances extending
from the BLR ($\sim 10^{-2}$ pc) to the NLR ($\sim 10^{3}$ pc).
Each cloud at radial distance $r$ from the nucleus represents the 
gas in the emission region described by the parameters:
hydrogen number density, $n_{\rm H}$ [cm$^{-3}$], dimensionless ionization 
parameter, $U$, total hydrogen column density, $N_{\rm H}$ [cm$^{-2}$], 
and the chemical abundances.  We note that, in the recent years, there is a 
growing evidence that the emitting and absorbing clouds are radiation pressure confined, 
and thus the total 
(radiation+ gas) pressure inside the cloud is constant with stratification in matter density
\citep{pier1995,dopita2002,groves2004a,groves2004b,rozanska2006,stern2014,baskin2014,adhikari2015,stern2016}. However, we have checked the assumption that each cloud is in pressure equilibrium does not change the main conclusion of our paper, therefore here we consider
the constant density clouds to keep the consistency with the NL93 formalism.
The ionization parameter is defined as the ratio of the number of hydrogen-ionizing 
photons, $Q_{\rm H}$ [s$^{-1}$], to the gas density of
the cloud \citep{osterbrock2006}
\begin{equation}
\label{eq:U}
U=\frac{Q_{\rm H}}{4\pi r^{2}n_{\rm H}~ c}
\end{equation}
where $c$ is the velocity of light. 

\begin{figure}[h]
\includegraphics[width=0.48\textwidth]{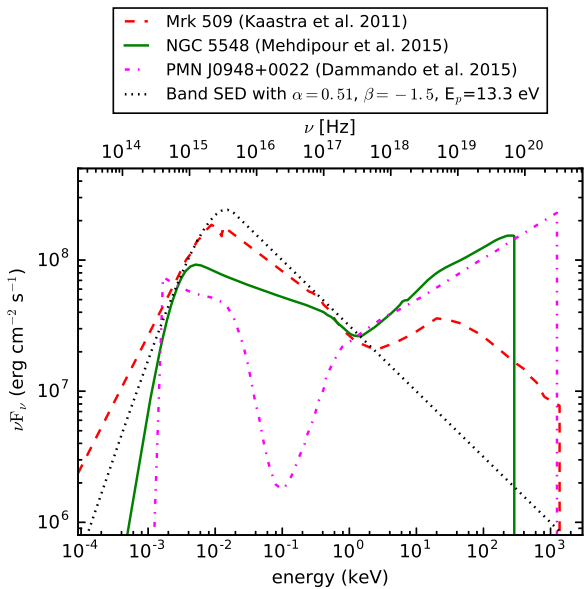}
\caption{\small Shapes of the broad band spectra used in our photoionization calculations. 
All SEDs are normalised to the
$ L = 10\rm {^{45}~erg~s^{-1}}$ and represent the radiation illuminating 
the cloud at $r$ = 0.1 pc. 
The dashed line shows Mrk 509, while the
solid line describes NGC 5548.  The SED of the
NLSy1  PMN J0948+0022 is presented as the
dashed dotted line, and the SED produced with the Band
function is shown by the dotted line.
See Section~\ref{sec:parameters} for parameters and references.} 
\label{fig:seds}
\end{figure}

These clouds are illuminated by the radiation of different spectral shapes 
shown in Fig.~\ref{fig:seds}. To consider various types of AGN we used the 
SED of the Sy1.5 galaxy Mrk 509 as measured in
multi-wavelength observation campaigns \citep{Kaastra2011} and 
the Sy1 galaxy NGC 5548 \citep{Mehdipour2015}. 
The SED of Mrk 509 is dominated by  soft X-ray photons below 1 keV, 
whereas the SED of NGC 5548 is dominated by harder photons above 2 keV.
To represent the SED of a NLSy1
we use the galaxy PMN J0948+0022 \citep{ammando2015}, shown as the magenta dashed line.
The NLSy1 galaxy PMN J0948+0022 SED has no pronounced emission around $\sim$ 0.1 keV,
rather it has an  excess of harder photons.
We also consider the spectral shape produced with the Band
function $f(E)$ \citep{band1993} as another typical  active galaxy.
Band function combines two power laws smoothly
and provides the possibility of creating  different shapes of spectra 
by allowing to change the parameters in the following expression:
\begin{equation}
\label{eq:bandf}
\begin{split}
f(E) & =  A\Big[\frac {E}{100}\Big]^{\alpha}~e^{\Big[\frac{-2(E+\alpha)}{E_{\rm p}}\Big]},~~~ {\rm for}~ E<\frac{(\alpha-\beta)E_{\rm p}}{(2+\beta)}\\
 & = A\Big[\frac{(\alpha-\beta)E_{\rm p}}{100(2+\alpha)}\Big]^{(\alpha-\beta)}\Big(\frac{E}{100}\Big)^{\beta}e^{(\beta-\alpha)},\\
 & ~~~~{\rm for}~ E\geq\frac{(\alpha-\beta)E_{\rm p}}{(2+\beta)}\\
\end{split}
\end{equation}
where $\alpha$ and $\beta$ are the slopes of the first and
second power law respectively. $E_{\rm p}$ is the peak energy, where the 
two power laws combine smoothly.
For producing the Band SED shape, we choose the slope
of the first power law $\alpha=0.51$ which cuts off
exponentially at $13.3$ eV and the second power law with slope $\beta=-1.5$.
These SEDs are chosen for two reasons:
(i) they are recently constrained by combining the data
from multi-wavelength observations (except the Band SED)
and (ii) they represent different possible SEDs that
an AGN can have in general. 

For deriving the luminosities
of the emission lines corresponding to each individual
portion of gas at the given distance, 
we make simulations with the photoionization {\sc cloudy} code (version  c13.03)
\citep{ferland2013}. 
We use the {\it luminosity case} in {\sc cloudy} with an assumed source luminosity of
$L=\rm~10^{45} ~ erg~s^{-1}$. Thus, the ionization parameter $U$
for each cloud is computed by the code itself.
The shape of the continuum is given by points using the
{\it interpolate} command (for details see {\sc cloudy}
documentation files\footnote{http://www.nublado.org/}).
We employed the profiles of the cloud parameters as
a function of $r$ as described by NL93:
\begin{equation}
\label{eq:params}
n_{\rm H}\propto r^{-3/2},~~~ N_{\rm H}\propto r^{-1}.
\end{equation}

To investigate  the effect of the SED on the resulting
line luminosities, the normalization of parameters in
Eq.~\ref{eq:params} is  chosen after NL93 that at $r=0.1$ pc,  
$n_{\rm H}=\rm ~10^{9.4} ~ cm^{-3}$ and $N_{\rm H}=\rm~10^{23.4}~cm^{-2}$.
With this choice of normalization, the values of $n_{\rm H}$ and
$N_{\rm H}$ of the cloud at the closest distance, $r$ = 10$^{-2}$ pc, are 
${\rm 10^{10.9} ~ cm^{-3}}$ and ${\rm 10^{24.4}~cm^{-2}}$ respectively.
The most distant cloud, at $r$ =10$^{3}$ pc, has values of  
${\rm 10^{3.4} ~ cm^{-3}}$ and  ${\rm 10^{19.4}~cm^{-2}}$ respectively. 
As justified in NL93,
these numbers are reasonable for simulating the BLR-NLR system of an
AGN with a bolometric luminosity of $10^{45}$~ erg~s$^{-1}$. 
The choice of the sublimation radius 
at $0.1$ pc is not obvious, since it depends on the size and the grains composition \citep{netzer2013}. 
However, for the purpose of understanding
the global properties of the line emission vs radius this 
assumption is reasonable and was also used by NL93.

We used the {\sc cloudy} default solar composition
for clouds located at  $r\leq$ 0.1 $\rm pc$ and for
more distant clouds at $r>$ 0.1 $\rm pc$
the interstellar medium (ISM) composition with
dust grains is used. This includes the graphite and silicate
component appropriate for the ISM of our galaxy 
and is fully taken into account by the {\sc cloudy} code. 

To investigate the role of the cloud density on our results, we change the normalization 
 to give different values of the gas density number at the sublimation radius. 
We consider the following four different number densities:
$n_{\rm H}=\rm~10^7, 10^{10}, 10^{11}, 10^{11.5} ~ cm^{-3}$ at $r=0.1$~pc 
while keeping the other parameters same.

\section{Line emission vs radius for various SEDs}
\label{sec:sed_dependence}

\begin{figure}
\hspace{-0.6cm}
\includegraphics[width=0.53\textwidth]{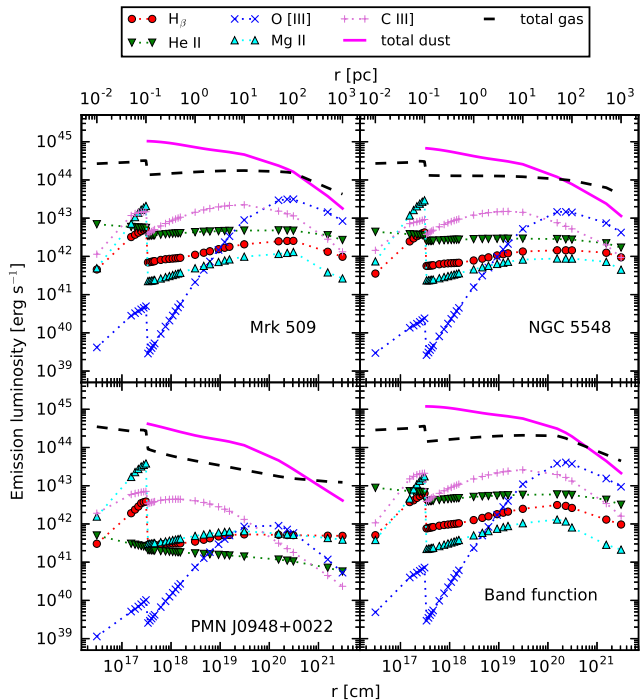}
\caption{\small The line emission vs radius for various lines:
H${\beta}$ (red circles), He~II (green triangles down), [O~III] (blue crosses), 
Mg~II (cyan triangles up) and C~III] (magenta pluses) shown as a function of radius
from the illuminating source. For all panels $n_{\rm H}=10^{9.4}$ cm$^{-3}$
at $r$ = 0.1 pc.  Total dust emission (magenta continuous line) and
the total gas emission (black dashed line) are shown for clarity.
Upper left panel presents the results for spectral shapes of Mrk~509, upper right panel 
- NGC 5548, lower left - NLSy1 PMN J0948+0022, and lower right panel 
- Band function.} 
\label{fig:line_profile}
\end{figure}

In this section we set the density normalization as in NL93, so that
$n_{\rm {H}}$=10$^{9.4}$ cm$^{-3}$ at $r$ = 0.1 pc.
The emission-line luminosity radial
profiles for the lines: H${\beta}$ (red circles), He~II
(green triangles down), Mg~II (cyan triangles up), C~III] (magenta pluses)
and [O~III] (blue crosses) are shown in
Fig.~\ref{fig:line_profile}. 
The upper panels of the figure show the line
luminosities for Mrk 509 (left) and NGC 5548 (right). In the
lower panels, results for PMN J0948+0022 (left) and Band function 
(right) are presented.

The nature of the line emission vs radius is similar for
Mrk 509, NGC 5548, and the Band function, independently of the shape
of the illuminating radiation. 
For all the  considered lines, the radial emission profiles 
display the strong suppression (by a factor of $\sim$ few 
to a few orders of magnitude)
of the line luminosity in a region
close to the sublimation radius. The suppression of the emission can 
also be noticed in the profile of the total gas emission
(black dashed lines in the figure). These results are in agreement with the
result of NL93 performed for only the one standard AGN SED \citet{Mathews1987} used by those authors.
However, the radial luminosity profile of the He~II line is only
weakly suppressed by the factor of $\sim$ 3
with the inclusion of dust. This is different from the case of NL93,
where the He~II emission goes down by $\sim$ an order of magnitude. 

Considerably
different radial emission profile
for the semi-forbidden line C~III] and forbidden line [O~III] is obtained in case 
of NLSy1 PMN J0948+0022 SED.
The C~III] line luminosity decreases steeply with increasing radius
showing a different behaviour from the other SEDs, where
it increases in a region between 0.1$<r<$10 pc.
As can be seen from the Fig.~\ref{fig:line_profile} (lower left panel),
the [O~III] emission is not significant in a region corresponding to the
NLR ($r~\sim$10~-~100 pc), which is different from the other SEDs
where an excess of [O~III] emission is seen at those distances.
Interestingly, this result is consistent with the SDSS spectrum of 
PMN J0948+0022, where  [O~III] emission is not present
\citep[][see their Fig.~2, lower panel]{tanaka2014}.
Finally, our model shows that He~II emission in this object is not
significantly affected by the presence of dust.  Rather, it is fairly constant
close to 0.1 pc and decreases slowly as a function of radial distance.
The differences in line emission of C~III], [O~III] and He~II 
from the other SEDs considered in our study are 
connected with the fact that the PMN J0948+0022 SED
has fewer photons in the extreme UV and soft X-ray band. This significantly smaller number
of hydrogen ionizing photons causes the drastic drop, by two orders of magnitude,
in lines emitted in the lower density dusty medium on the larger radii.

Despite these differences, the effect of dust on the line suppression is 
clearly visible in the case of PMN J0948+0022, and it is the strongest 
for the high luminosity H${\beta}$ and Mg~II lines. 
The general conclusion is that for  gas with physical
conditions and parameters 
given by NL93, the apparent gap in the line emission cannot be removed 
by changing the shape of illuminating continuum. 
The differences in the [O~III], C~III] and He~II radial luminosity profiles
for NLSy1 SED are not sufficient to explain the origin of the
ILR. 

\section{Role of the gas density}
\label{sec:density_dependence}

\begin{figure}
\hspace{-0.6cm}
\includegraphics[width=0.53\textwidth]{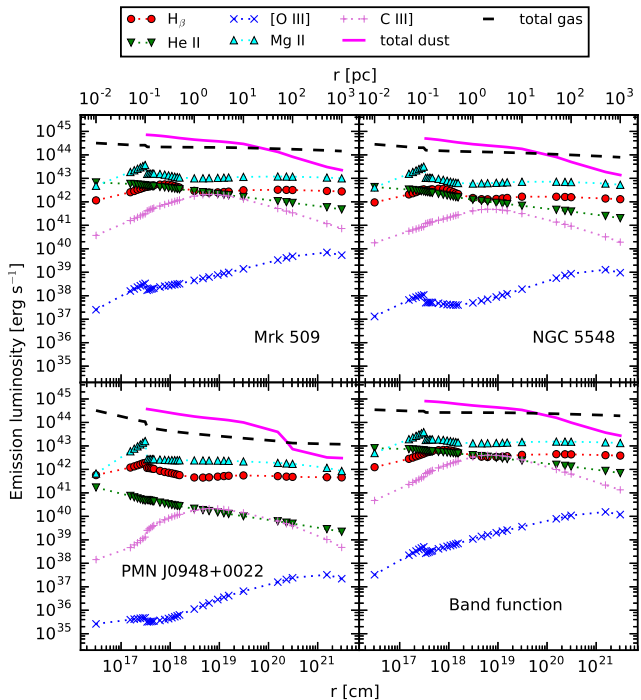}
\caption{\small 
Same as in Fig.~\ref{fig:line_profile}, but density $n_{\rm H}=\rm ~10^{11.5} ~ cm^{-3}$ at the sublimation radius, $r$=0.1 pc.} 
\label{fig:line_profile_hd}
\end{figure}

High local densities in the BLR ($n_{\rm H}$ $\sim 10^{11} - 10^{11.5}$ cm$^{-3}$)
have been deduced from the study of  UV Fe~II emission of an extreme NLSy1 object, I Zw 1
\citep{bruhweiler2008}, and in the quasars:  LBQS 2113-4538 \citep{hryniewicz2014}, CTS C30.10
\citep{modzelewska2014}, and HE 0435-4312 \citep{sredzinska2016}.
Additionally, a density  of $10^{10.25}$ cm$^{-3}$ was found from the study of UV 
intrinsic absorbers of two NLSy1 galaxies  \citep{karen2004}. 
Finally, from photoionization calculations, \citet{rozanska2014} found the density of 
an intrinsic absorbing cloud in the bright quasar HS 1603 + 3820 to be
of the order of $n_{\rm H}$ $\sim 10^{10} - 10^{12}$ cm$^{-3}$.
On the basis of those results, we increased the normalization of the gas density 
at  $r$=0.1 pc, to be $n_{\rm H}=\rm~10^{11.5} ~ cm^{-3}$ in our model. 
Note, that the ionization parameter changes inversely with $n_{H}$
as given in Eq.~\ref{eq:U}, but all other parameters have the same values 
as described in Sec.~\ref{sec:parameters}. 

The resulting emission profiles of the five major lines are shown
in Fig.~\ref{fig:line_profile_hd} for the  four SEDs. 
The meaning of lines and points is exactly the same as in Fig.~\ref{fig:line_profile}.
The H${\beta}$, 
He~II, and C~III] emission lines behave differently from the 
 lower density normalization case. 
The presence of dust does not suppress  H${\beta}$ and C~III]
line emission, rather it is enhanced by the factor of
$\sim$ 1.5 at $0.1<r<0.3$ pc, in the case of the Mrk~509, NGC~5548 and  Band
function SEDs. For the SED of NLSy1 PMN J0948+0022 C~III] emission increases
by the factor of $\sim$ 3 at $0.1<r<0.3$ pc, whereas H${\beta}$ emission
decreases slightly with the distance, but there is no large jump 
at the sublimation radius, as was found in the lower density
case. Only the Mg~II line displays a jump at 0.1~pc, but it is much less prominent than in 
the lower density case. In all other cases we observe the lack of the line suppression 
for all shapes of SED.

Our result clearly shows that the strong 
suppression in the line emission  at the sublimation radius 
for lower density normalization, disappears at the higher density.
Therefore, intermediate line emission can be produced at radial distances 
$r~\sim$ 0.1 - 1 pc. Note, that the gas emission 
also does not display any jump for high density case. 
The strong ILR emission found by
\citep{mason1996,crenshaw2007,crenshaw2009}, can be produced 
if the clouds are two orders of magnitude denser than those 
considered by NL93. Our results corroborate with the speculation made
by \citet{landt2014}, where they discuss the possible role of high density gas for the production of smooth BLR.
Furthermore, if the ILR is located at distances predicted by our model, we can estimate 
the expected reverberation mapping lag. Within the framework of our model, the ILR lag would be of the order of 100-1000 light-days. 

\begin{figure}
\hspace{-0.6cm}
\includegraphics[width=0.53\textwidth]{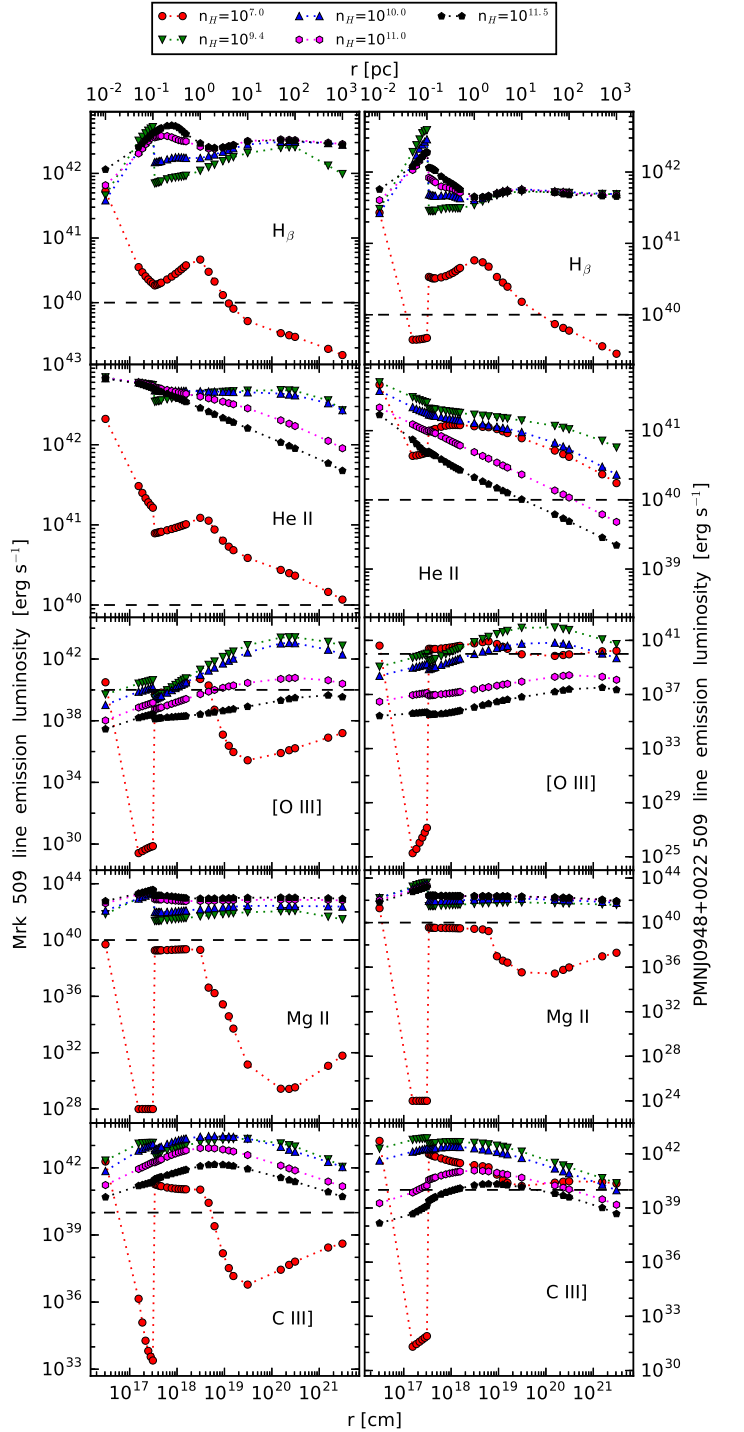}
\caption{\small 
The line emission vs radius for different densities:
$ n_{\rm H}=10^{7.0}$~(red~circles),~ $10^{9.4}$~(green~triangles~down),
$10^{10.0}$~(blue~triangles~up),~ $10^{11.0}$~(magenta~hexagons)
and ~ $10^{11.5}$ (black~pentagons) cm$^{-3}$ at $r=0.1$ pc. 
The left and right panels show the case of the Sy1.5 Mrk 509 
and NLSy1 PMN J0948+0022 SEDs respectively.
Five major lines: H${\beta}$, He~II, [O~III], Mg~II and C~III] are presented from top to  
bottom. The detection limit on line luminosity is 
shown by horizontal black dashed line (see text for details).}
\label{fig:varden}
\end{figure}

The physical reason for the density dependence is connected with the value of the column density 
at which the ionization front is located. Below, we explain this in the example of hydrogen 
H${\beta}$ line. Let us assume simple ground-state hydrogen photoionization, and set the ionization balance equation
\begin{equation}
n_{\rm p}~n_{\rm e}~ l~ \alpha_{\rm B}(T) = \phi_{\rm H}.
\end{equation}
\citep{osterbrock2006}.
The first two terms are the proton and electron density in the H$^+$ layer of the cloud,
$l$~[cm] is the thickness of the H$^+$ layer, and $\alpha_{\rm B}(T)$~[cm$^{3}$~s$^{-1}$] is the 
Case B recombination coefficient. The physical interpretation is that the flux of hydrogen-ionizing 
photons incident on the cloud, $\phi_{\rm H}$~[cm$^{-2}$~s$^{-1}$], equals the number of 
hydrogen recombinations that occur over the thickness $l$.

\begin{figure}
\hspace{-0.6cm}
\includegraphics[width=0.56\textwidth]{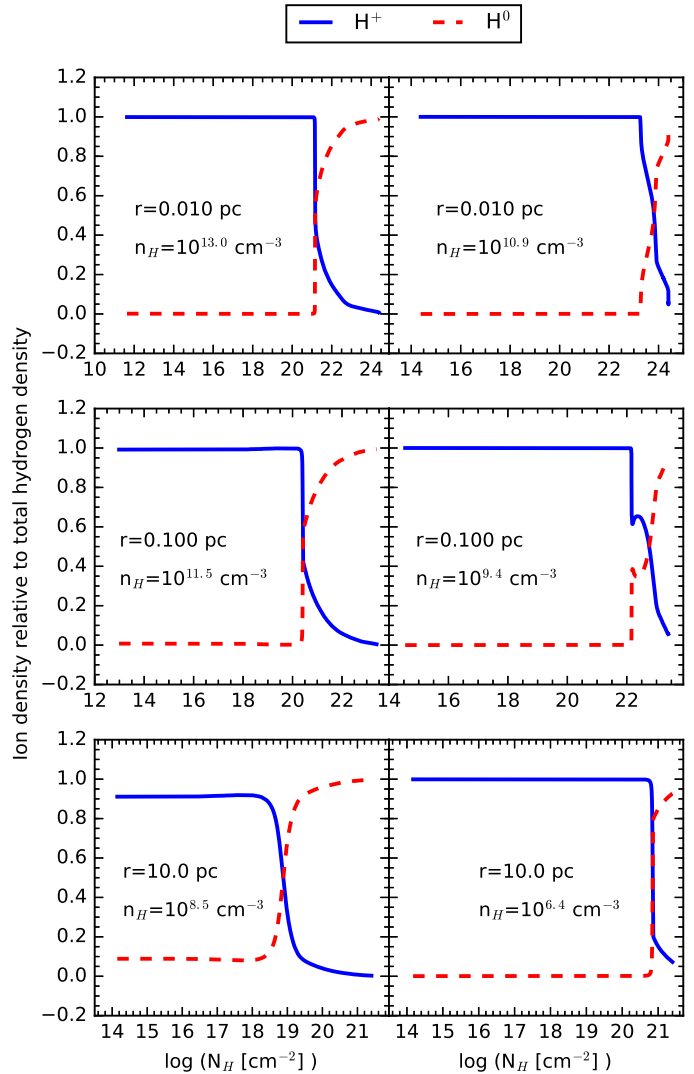}
\caption{\small Ionized (H$^{+}$) and neutral (H$^{0}$) hydrogen densities 
relative to the total cloud denesity $n_{\rm H}$, as a function of  
column density across the cloud  illuminated by SED of Mrk 509. 
The left and right panels show the ion densities for $n_{\rm H}= 10^{11.5}$ and $10^{9.4}$
cm$^{-3}$ at $r = 0.1$ pc respectively. Values of $n_{\rm H}$ change
with $r$ as Eq.~\ref{eq:params}, and are given in each panel together with the
cloud location.}
\label{fig:ion_density}
\end{figure}

The hydrogen column density across H$^+$ layer is then
\begin{equation}
\label{eq:h+col}
N_{\rm H^+} \equiv n_p l =  \frac{\phi_{\rm H}}{n_{\rm e} \alpha_{\rm B}(T)} = U c~\alpha_{\rm B}(T)^{-1} .
\end{equation}
The gas column density $N_{\rm H^+}$, and resulting line and continuum optical depths, 
all scale with the ionization parameter. At very high values of $U$, whole cloud is ionized, 
and the line emission comes from the whole volume, in this case limited 
by the fixed adopted total hydrogen column, $N_{\rm H}$.
As the cloud density increases, the ionization 
parameter decreases, ionization front forms, and the cloud consist of two zones. 
The line emission comes from the first zone, and only the dust in this zone competes 
with the gas for the photons. The high-density clouds
studied here have smaller column densities and optical
depths of H$^+$ layer, since ionization parameter is lower.
This is clearly demonstrated in Fig.~\ref{fig:ion_density}, which shows that
the H$^+$ - H$^0$ ionization front moves towards smaller cloud
thickness with increasing number density of the cloud. 
For smaller values of $U$, the dust absorption decreases,
so there is less suppression of the emission due to the
presence of dust. In other words, the less dense clouds
have much larger thickness of H+ layer, larger dust column density in the 
zone with abundant photons, and therefore the
dust absorption is significant.

Eq.~\ref{eq:h+col} can be used for the quantitative estimation of threshold U,
below which the gas opacity for incident photons dominates over opacity of dust. 
Taking recombination coefficient $\alpha_{\rm B}~(\rm at ~10^{4}~ K)~= $~ 2.6$\times 10^{-13}$ cm$^{3}$~s$^{-1}$, gives the column density:
$N_{\rm H^+}~\sim~$ 10$^{23}~U$. Therefore, the dust optical depth for UV photons is 
$\tau_{\rm dust} \cong N_{\rm H^+}/10^{21}$ = 100 $U$ \citep[for details see][]{guver2009}. 
For $U$=0.1,
$\tau_{\rm dust}$ =10 at the edge of the H$^{+}$ region. 
This implies that the width of the H$^{+}$ layer where gas actively 
absorbs and emits is only 1/10 of the 
total layer. 
In other words, increasing the density by the 
factor $>$ 10 would mean $U~<$ 0.01 which implies $\tau_{\rm dust}$ $<$ 1. 
This means that the gas opacity always dominates for higher densities  and 
it does not matter if the gas is dusty or not, and therefore no
suppression of the line emission is physically possible. 

 It is worth to mention the conclusion of
\citet{ferland1983}, saying that for low ionization nuclear emission line region (LINERs), if 
the line emission comes from the photoionised gas, the  
 $U$ is $\leq~\rm 10^{-3}$. This is less than the threshold $U$
corresponding to higher density (10$^{11.5}$ cm$^{-3}$) implied by our result,
where the gas opacity is always dominant over the opacity of dust. So, our result 
clearly indicates that LINERs should also exhibit the ILR. We note that the 
presence of ILR in LINERs is also discussed by \citet{balmaverde2016} 
where they analysed 33 LINERs (bona fide AGN) and Seyfert galaxies from 
optical  spectroscopic  Palomar  survey  observed  by  {\it HST/STIS}. 
However, the density of the outer portions of ILR claimed by those authors is about three orders 
of magnitude less (i.e., 10$^4$-10$^5$ cm$^{-3}$) than what our model predicts. Most probably this 
is due to the fact that LINERs are much fainter than objects considered in this paper.

\section{Discussion}
\label{sec:disc}

\subsection{Individual line behaviour}

Beside the general trend presented
above, in some cases we note exceptional line behaviour. 
For instance, Mg~II (cyan triangles up in Figs.~\ref{fig:line_profile}, 
and \ref{fig:line_profile_hd}) line emissivity does not depend 
on the shape of the SED. All SEDs considered by us
produce the emissivity jump at the sublimation radius, but this jump is  
two orders of magnitude higher when the cloud  density is lower,
due to physical reason given above, but for the case of magnesium ionization front.  

The density influence on the suppression of line luminosity jump at the sublimation 
radius is best seen in case of H${\beta}$. For all continuum shapes the 
strong jump in emission (order of magnitude) is present for the canonical density 
used by NL93, $n_{\rm H}=10^{9.4}$~cm$^{-3}$. This jump is not seen at all
for much denser clouds. 

Note, that the dust influence on the emission jump is much stronger in case 
of Mg~II line than for H${\beta}$ line. 
This difference is due to 
the fact that the H${\beta}$ line forms in deeper parts of a cloud than Mg~II line. 
In the presence of dust, the radiation reaching these depths is 
harder since the dust opacity selectively removes lower energy photons, 
allowing harder photons to reach the 
region where H${\beta}$ is formed. 
As a result, the H${\beta}$ forming region is warmer and 
more ionized making the line stronger. Moreover, the 
dust opacity is smaller at the wavelength 
of H${\beta}$ than Mg~II, so H${\beta}$ is absorbed 
less than Mg~II. This implies that the contribution of the NLR to the
overall line shape is expected to be lower in Mg~II than in H${\beta}$.

The He~II line emission vs radius from our simulations shows the luminosity jump 
is at least one order of magnitude lower that the jump reported by NL93 for the same line. 
Our simulations show that this jump is much smaller, only a factor of two, for three Sy1 SEDs, 
and completely disappears in the case of the NLSy1 continuum.
Those differences may be caused by the different numerical code used by those authors. 
Nevertheless, with the high density case, He~II line suppression is not 
present for all considered  shapes of radiation. 

For both densities, the C~III] line strongly depends on the SED shape. 
With the   NLSy1 PMN J0948+0022 SED and the dust presence, line luminosity slowly decreases 
with distance by $\sim$few factors, and the maximum line emission always occurs at $r<0.1$~pc.
In case of the Sy1 SEDs, the line luminosity peaks at $r \sim 10$~pc, with clear suppression at the
sublimation radius for the low-density case. 

The  [O~III] emission (blue crosses in Figs.~\ref{fig:line_profile}, 
and \ref{fig:line_profile_hd}) reaches a maximum  farther outside the
sublimation radius. For the high density case, 
this line is very weak in comparison with other lines. This is in agreement with 
the fact that forbidden lines originate from low density gas. 
Only for low density case there is  enhanced 
[O~III] emission in a region close to the NLR ($r~\sim$10~--~100 pc). We expect that such line 
will never be observed in the ILR. Indeed, the initial claim of the [O~III] variability
was not supported by the careful analysis of the data \citep{barth2016}.
Beside forbidden [O~III] line, our results strongly support the presence of an ILR 
 where the intermediate velocity lines are expected. 

\subsection{Apparent gap suppression}

To investigate  the role of  density in the formation of the ILR, we calculated 
our model for several values of density normalizations:
$n_{\rm H}=10^{7.0},~ 10^{9.4},~ 10^{10.0}, ~ 10^{11.0},~ 10^{11.5}$~cm$^{-3}$,
keeping other parameters the same. 
Fig.~\ref{fig:varden} shows the dependence of
line luminosity radial profiles on the density normalization for two representative 
SEDs: Mrk 509 (left panels)
and PMN J0948+0022 (right panels), for all lines considered by us.

For the rare clouds, the luminosity of some lines decreases many orders of magnitude. 
Those lines would
not be observed due to the sensitivity of the telescopes.
Therefore, we mark the lower limit on the line luminosity above which line could be visible, 
by the horizontal dashed line on each panel. We estimated this limit by assuming a Gaussian
 spectral profile  of the emission lines
on top of underlying continuum given by our SEDs. For
this calculation, spectral line width is set to the value corresponding 
to Keplerian velocity at 1 pc ($\sim$ ILR radii), for a
BH mass $\sim 10^{8} M_{\odot}$.
It is rather hard lower limit on the observable 
line luminosities since the simulated isotropic luminosities should be at least 10 times 
higher (assuming 100\% covering factor).
Nevertheless, simulated line luminosities presented in Fig.~\ref{fig:varden} fall below this 
limit in some cases, and those lines have no chance to be
detected.

Note, that for a given SED, there is a preferred value of density for which the line 
emission is the greatest. For example; the H${\beta}$ luminosity increases with the density
normalization and is a maximum for $n{\rm_{H}=10^{11.5}~cm^{-3}}$  at distances 
0.1 pc $< r < 0.4$ pc. The He II luminosity  peaks at $r < 0.1$ pc, and the maximum 
emission occurs for the lower density, whose value depends on the SED shape. 
In case of PMN J0948+0022 the luminosity peak
is lower by an order of magnitude than for Mrk 509
SED, and occurs for density $n_{\rm {H}}{\rm ~=~10^{7} ~cm^{-3}}$. This is in
agreement with Locally Optimised Cloud (LOC) model
of \citet{baldwin1995}.

The radial distances at which the H${\beta}$ and He~II emissions are the strongest are 
in agreement with the results inferred from the reverberation studies of the BLR in AGN. 
The radial stratification with ionization potential of the species producing the line has been 
observed \citep[i.e.][]{clavel1991,peterson1999}, showing
that He~II line always originates at distances closer by a
factor of three/four to the nucleus than H line.
In addition, the distances for which the luminosities of Mg~II
and [O~III] peaks are consistent with the reverberation
mapping studies. 

For the Mrk 509 SED, the emission, with density normalization 
$n_{\rm {H}}{\rm ~=~10^{7} ~cm^{-3}}$ at 0.1 pc, is insignificant for
all considered lines. However, the [O~III] emission is the
strongest for the PMN J0948+0022 SED with the lowest
density , which agrees with the prediction that forbidden
lines are formed in low density gas. For both SEDs, the
semi-forbidden line C~III] has a maximum emission at
intermediate densities i.e, $n_{\rm {H}}{\rm ~=~10^{9.4} ~cm^{-3}}$ at 0.1 pc.

Despite of the different distances where the line luminosities peak the strong jump 
in the emission profiles disappears in almost all cases for both shapes of continuum, 
when density normalization increases. The lack of this jump clearly means that there will 
not be a gap between the BLR and NLR. This naturally explains the
origin of the intermediate line region.

\subsection{The connection with an accretion disk}

\begin{figure}
\hspace{-0.4cm}
\includegraphics[width=0.54\textwidth]{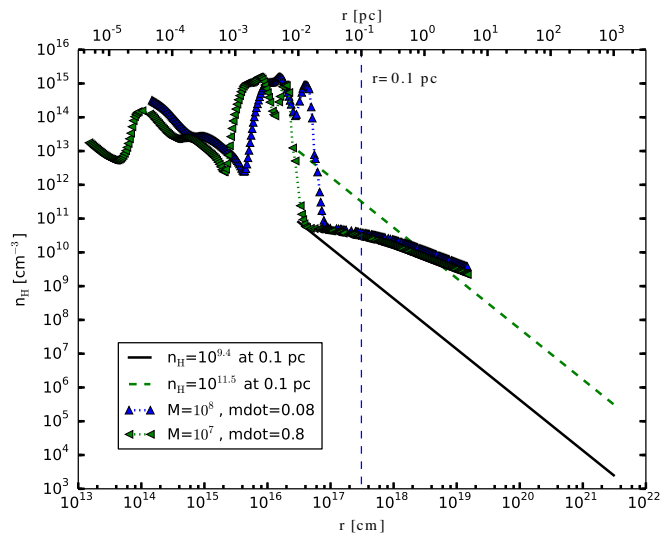}
\caption{\small The accretion disk atmosphere density radial profiles. 
The vertical dashed line marks the position of the sublimation radius. Straight lines are 
plotted according to the power-law given in Eq.~\ref{eq:params} by the first term for
two different density normalization marked at the figure. Two datasets are 
presented for expected masses and 
Eddington ratios derived from an assumed bolometric 
luminosity ($10^{45}$ erg~s$^{-1}$, see text for details).}
\label{fig:disk}
\end{figure}

It is widely believed that broad line emission clouds may be connected 
with the wind from an accretion disk atmosphere 
\citep{gaskell2009,czerny2011}. 
The upper atmospheric layers of the disk can be quite dense, with value up to 
$n_{\rm H} \sim 10^{14}$~cm$^{-3}$, depending on the distance from the central black hole
\citep{hrynio2011,rozanska2014}.  These densities are derived
by solving for the accretion disk vertical structure in hydrostatic and radiative equilibrium 
parametrized by the black hole
mass, its spin, and the accretion rate \citep{rozanska99}. 
We can properly derive the density at the disk photosphere, i.e. at the optical 
depth  $\tau \sim 2/3$, assuming energy generation via viscosity, and diffusion 
approximation of the radiative transfer. 
Our calculations use the  Rosseland mean opacity tables from 
\citet{alexander83,seaton94}. Those opacities are crucial for calculating the disk 
density  since the true absorption opacity can be an order of magnitude
higher than electron scattering opacity as shown by 
\citet[][see their Fig.~4 and 7]{rozanska99}.

We take the density at $\tau=2/3 $ as a representative of the initial 
density in the disk wind, which supplies matter to the BLR or intrinsic absorbers in an outflow. 
The model fully takes into account the radial
 transition from pressure dominated regions to gas dominated regions by assuming that the
viscous torque is proportional to the total pressure (gas and radiation), and we use realistic 
description of the opacities, including atomic transitions and the presence of the dust. Nevertheless, 
the model should not be extended too far beyond a $\sim 1$ pc scale because there the self-gravity 
effects as well as the effects of the circumnuclear stellar cluster become important
\citep{thompson2005}.

The disk atmosphere radial density profile at $\tau=2/3$ is plotted in Fig.~\ref{fig:disk}.
We used two test sets of parameters. Keeping the bolometric luminosity constant 
at $10^{45}$~erg~s$^{-1}$ we find the Eddington ratio for 
masses of $10^7 M_{\odot}$ and $10^8 M_{\odot}$. 
We can assume that our rescaled  SEDs should be produced in AGNs with BH masses in this range.
For the given assumptions and taking a radiative efficiency of 0.1 we then derive the Eddington 
ratios corresponding to those masses: 0.8  and 0.08 respectively. In addition, the density of 
the clouds used in our BLR-NLR simulations is drawn by straight lines according to
Eq.~\ref{eq:params}, for two normalizations at the sublimation radius:
$n_{\rm H} = 10^{9.4}$, and $10^{11.5}$ cm$^{-3}$.

It is thus clearly seen, that accretion disk atmospheres
naturally produce high densities at the sublimation radius. All models show a strong change in 
density produced by heavy element Rosseland mean opacities used in our calculations. 
Even if this bump
is slightly inside sublimation radius assumed by us, the
density radial profiles show that high densities occur in
the disk atmosphere, and such dense gas can give rise to the line emitting regions.
 
On the other hand, the local density at the disk atmosphere does not depend 
on the Eddington ratio, and is expected to be the same 
for the Sy1 and NLSy1 galaxies. Thus the differentiation in the local density 
between those two types of objects must happen at the wind formation stage. 
Smooth wind outflow causes the decrease of the density with the distance measured 
along the wind stream line due to wind acceleration and the geometrical divergence 
of the wind line. However, if thermal instability operates in the wind and colder/denser 
cloud forms the evolution of the wind density becomes very complex and dependent on the 
heating and cooling efficiency. The difference between the wind developing in Sy1 and NLSy1 
may be due to the difference in the basic wind velocity. A given disk radius, measured in cm 
or pc, corresponds to larger value of the $r/R_{\rm Schw}$ in NLSy1 than in Sy1, the corresponding 
Keplerian velocity is lower, and the wind velocity, which usually is of the same order, is also 
smaller, leaving perhaps more time for the development of the thermal instability in the wind. 
This would explain higher density of the clouds in NLSy1 than in Sy1. However, detailed 
analysis of the time-dependent wind model is certainly beyond the scope of the current paper.

\section{Conclusions}
\label{sec:con}
Following the work of NL93, we performed  
numerical simulations of photoionised gas in 
AGN  emission line regions and derived the
line luminosity radial profiles for various major lines:
H${\beta}$ ${\lambda}$4861.36 \AA, He~II ${\lambda}$1640.00 \AA,
Mg~II ${\lambda}$2798.0 \AA, C~III] ${\lambda}$1909.00 \AA
~and [O~III] ${\lambda}$5006.84 \AA ~originating at different 
distances from the central engine of the AGN. 

On the basis of the line emission vs radius derived for four different 
SED shapes of incident radiation, and for different density normalizations at the sublimation
 radius, we find the following:
\begin{enumerate}
\item The presence or absence of the Intermediate Line Region is not determined by the spectral 
shape of the incident continuum. Different SEDs do produce considerably different 
behaviour of the emission line radial profiles, due to the different amount of 
extreme UV and soft X-ray photons in its broad band SED. However these differences
do not adequately explain why an intermediate line emission is observed in some objects.
\item With higher densities normalization, i.e., n$_{H}$=10$^{11.5}~\rm{cm^{-3}}$ at $r=0.1$~pc,
we obtained flat luminosity line radial profiles for almost all lines and SEDs.
Thus, the dust does not suppress the line emission, contrary to the result obtained
by NL93. Our model potentially explains the existence of ILR in some sources. If the density 
of the gas is high enough, emission lines of intermediate velocity width can be produced.
\item Such ILR is predicted to be located at radial distances $r \sim 0.1-1$ pc, and 
the expected by our model the reverberation mapping lag would be of the order of 100-1000 light-days.
\item We demonstrated the dependence of line luminosity profiles on the density normalization. 
We found that the significant line emission in objects with a particular SED occurs at different 
densities in agreement with reverberation mapping studies of H${\beta}$, Mg~II, [O~III] and He~II
lines. 
\item  We showed that dense clouds postulated by us can be potentially formed from an accretion disk 
atmosphere which is dense enough below the sublimation radius in the accretion disk. This work is
another proof that the high densities occur in the BLR, which is in full agreement with many previous studies.
\end{enumerate}

\acknowledgments
{\small 
We are grateful to Jonathan Stern, the referee, for very helpful
comments to the manuscript, and we thank Ari Laor and Jian-Min Wang
for helpful discussion. This research was supported by Polish National Science 
Center grants No. 2011/03/B/ST9/03281, 2015/17/B/ST9/03422, and 
by Ministry of Science and Higher Education grant W30/7.PR/2013.
It received funding from the European Union Seventh Framework Program 
(FP7/2007-2013) under the grant agreement No.312789.
T.P.A. received funding from NCAC PAS grant for
PhD students. BC acknowledges the grant
from Foundation for Polish Science through the Master/Mistrz program 3/2012.
GJF thanks the Nicolaus Copernicus Astronomical Center for its hospitality 
and acknowledges support by NSF (1108928, 1109061, and 1412155), 
NASA (10-ATP10-0053, 10-ADAP10-0073, NNX12AH73G, and ATP13-0153), 
and STScI (HST-AR- 13245, GO-12560, HST-GO-12309, GO-13310.002-A, 
HST-AR-13914, and HST-AR-14286.001).
}



\bibliographystyle{apj}
\bibliography{refs}

\begin{thebibliography}{64}
\expandafter\ifx\csname natexlab\endcsname\relax\def\natexlab#1{#1}\fi

\bibitem[{{Adhikari} {et~al}\mbox{.}(2015){Adhikari}, {R{\'o}{\.z}a{\'n}ska},
  {Sobolewska}, \& {Czerny}}]{adhikari2015}
{Adhikari} T.~P., {R{\'o}{\.z}a{\'n}ska} A., {Sobolewska} M., {Czerny} B.,
  2015, \apj, 815, 83

\bibitem[{{Alexander}, {Rypma} \& {Johnson}(1983){Alexander}, {Rypma}, \&
  {Johnson}}]{alexander83}
{Alexander} D.~R., {Rypma} R.~L., {Johnson} H.~R., 1983, \apj, 272, 773

\bibitem[{{Baldwin} {et~al}\mbox{.}(1995){Baldwin}, {Ferland}, {Korista}, \&
  {Verner}}]{baldwin1995}
{Baldwin} J., {Ferland} G., {Korista} K., {Verner} D., 1995, \apjl, 455, L119

\bibitem[{{Baldwin} {et~al}\mbox{.}(1991){Baldwin}, {Ferland}, {Martin},
  {Corbin}, {Cota}, {Peterson}, \& {Slettebak}}]{baldwin1991}
{Baldwin} J.~A., {Ferland} G.~J., {Martin} P.~G., {Corbin} M.~R., {Cota} S.~A.,
  {Peterson} B.~M., {Slettebak} A., 1991, \apj, 374, 580

\bibitem[{{Balmaverde} {et~al}\mbox{.}(2016){Balmaverde}, {Capetti}, {Moisio},
  {Baldi}, \& {Marconi}}]{balmaverde2016}
{Balmaverde} B., {Capetti} A., {Moisio} D., {Baldi} R.~D., {Marconi} A., 2016,
  \aap, 586, A48

\bibitem[{{Band} {et~al}\mbox{.}(1993){Band}, {Matteson}, {Ford}, {Schaefer},
  {Palmer}, {Teegarden}, {Cline}, {Briggs}, {Paciesas}, {Pendleton}, {Fishman},
  {Kouveliotou}, {Meegan}, {Wilson}, \& {Lestrade}}]{band1993}
{Band} D. {et~al.}, 1993, \apj, 413, 281

\bibitem[{{Barth} \& {Bentz}(2016)}]{barth2016}
{Barth} A.~J., {Bentz} M.~C., 2016, \mnras, 458, L109

\bibitem[{{Baskin}, {Laor} \& {Stern}(2014){Baskin}, {Laor}, \&
  {Stern}}]{baskin2014}
{Baskin} A., {Laor} A., {Stern} J., 2014, \mnras, 438, 604

\bibitem[{{Boroson} \& {Green}(1992)}]{boroson1992}
{Boroson} T.~A., {Green} R.~F., 1992, \apjs, 80, 109

\bibitem[{{Brotherton} {et~al}\mbox{.}(1994){Brotherton}, {Wills}, {Francis},
  \& {Steidel}}]{brotherton1994}
{Brotherton} M.~S., {Wills} B.~J., {Francis} P.~J., {Steidel} C.~C., 1994,
  \apj, 430, 495

\bibitem[{{Bruhweiler} \& {Verner}(2008)}]{bruhweiler2008}
{Bruhweiler} F., {Verner} E., 2008, \apj, 675, 83

\bibitem[{{Clavel} {et~al}\mbox{.}(1991){Clavel}, {Reichert}, {Alloin},
  {Crenshaw}, {Kriss}, {Krolik}, {Malkan}, {Netzer}, {Peterson}, {Wamsteker},
  {Altamore}, {Baribaud}, {Barr}, {Beck}, {Binette}, {Bromage}, {Brosch},
  {Diaz}, {Filippenko}, {Fricke}, {Gaskell}, {Giommi}, {Glass}, {Gondhalekar},
  {Hackney}, {Halpern}, {Hutter}, {Joersaeter}, {Kinney}, {Kollatschny},
  {Koratkar}, {Korista}, {Laor}, {Lasota}, {Leibowitz}, {Maoz}, {Martin},
  {Mazeh}, {Meurs}, {Nair}, {O'Brien}, {Pelat}, {Perez}, {Perola}, {Ptak},
  {Rodriguez-Pascual}, {Rosenblatt}, {Sadun}, {Santos-Lleo}, {Shaw}, {Smith},
  {Stirpe}, {Stoner}, {Sun}, {Ulrich}, {van Groningen}, \&
  {Zheng}}]{clavel1991}
{Clavel} J. {et~al.}, 1991, \apj, 366, 64

\bibitem[{{Crenshaw} \& {Kraemer}(2007)}]{crenshaw2007}
{Crenshaw} D.~M., {Kraemer} S.~B., 2007, \apj, 659, 250

\bibitem[{{Crenshaw} {et~al}\mbox{.}(2009){Crenshaw}, {Kraemer}, {Schmitt},
  {Kaastra}, {Arav}, {Gabel}, \& {Korista}}]{crenshaw2009}
{Crenshaw} D.~M., {Kraemer} S.~B., {Schmitt} H.~R., {Kaastra} J.~S., {Arav} N.,
  {Gabel} J.~R., {Korista} K.~T., 2009, \apj, 698, 281

\bibitem[{{Czerny} \& {Hryniewicz}(2011)}]{czerny2011}
{Czerny} B., {Hryniewicz} K., 2011, \aap, 525, L8

\bibitem[{{Czerny} {et~al}\mbox{.}(2015){Czerny}, {Modzelewska}, {Petrogalli},
  {Pych}, {Adhikari}, {{\.Z}ycki}, {Hryniewicz}, {Krupa}, {{\'S}wie{\c
  t}o{\'n}}, \& {Niko{\l}ajuk}}]{czerny2015}
{Czerny} B. {et~al.}, 2015, Advances in Space Research, 55, 1806

\bibitem[{{D'Ammando} {et~al}\mbox{.}(2015){D'Ammando}, {Orienti}, {Finke},
  {Raiteri}, {Hovatta}, {Larsson}, {Max-Moerbeck}, {Perkins}, {Readhead},
  {Richards}, {Beilicke}, {Benbow}, {Berger}, {Bird}, {Bugaev}, {Cardenzana},
  {Cerruti}, {Chen}, {Ciupik}, {Dickinson}, {Eisch}, {Errando}, {Falcone},
  {Finley}, {Fleischhack}, {Fortin}, {Fortson}, {Furniss}, {Gerard},
  {Gillanders}, {Griffiths}, {Grube}, {Gyuk}, {H{\aa}kansson}, {Holder},
  {Humensky}, {Kar}, {Kertzman}, {Khassen}, {Kieda}, {Krennrich}, {Kumar},
  {Lang}, {Maier}, {McCann}, {Meagher}, {Moriarty}, {Mukherjee}, {Nieto}, {de
  Bhr{\'o}ithe}, {Ong}, {Otte}, {Pohl}, {Popkow}, {Prokoph}, {Pueschel},
  {Quinn}, {Ragan}, {Reynolds}, {Richards}, {Roache}, {Rousselle}, {Santander},
  {Sembroski}, {Smith}, {Staszak}, {Telezhinsky}, {Tucci}, {Tyler}, {Varlotta},
  {Vassiliev}, {Wakely}, {Weinstein}, {Welsing}, {Williams}, \&
  {Zitzer}}]{ammando2015}
{D'Ammando} F. {et~al.}, 2015, \mnras, 446, 2456

\bibitem[{{Davidson}(1972)}]{davidson1972}
{Davidson} K., 1972, \apj, 171, 213

\bibitem[{{De Robertis} \& {Osterbrock}(1984)}]{derobertis1984}
{De Robertis} M.~M., {Osterbrock} D.~E., 1984, \apj, 286, 171

\bibitem[{{Dopita} {et~al}\mbox{.}(2002){Dopita}, {Groves}, {Sutherland},
  {Binette}, \& {Cecil}}]{dopita2002}
{Dopita} M.~A., {Groves} B.~A., {Sutherland} R.~S., {Binette} L., {Cecil} G.,
  2002, \apj, 572, 753

\bibitem[{{Ferland} \& {Netzer}(1979)}]{ferland1979}
{Ferland} G., {Netzer} H., 1979, \apj, 229, 274

\bibitem[{{Ferland} \& {Netzer}(1983)}]{ferland1983}
{Ferland} G.~J., {Netzer} H., 1983, \apj, 264, 105

\bibitem[{{Ferland} \& {Osterbrock}(1986)}]{ferland1986}
{Ferland} G.~J., {Osterbrock} D.~E., 1986, \apj, 300, 658

\bibitem[{{Ferland} {et~al}\mbox{.}(2013){Ferland}, {Porter}, {van Hoof},
  {Williams}, {Abel}, {Lykins}, {Shaw}, {Henney}, \& {Stancil}}]{ferland2013}
{Ferland} G.~J. {et~al.}, 2013, \rmxaa, 49, 137

\bibitem[{{Gaskell}(2009)}]{gaskell2009}
{Gaskell} C.~M., 2009, \nar, 53, 140

\bibitem[{{Gaskell}, {Shields} \& {Wampler}(1981){Gaskell}, {Shields}, \&
  {Wampler}}]{gaskell1981}
{Gaskell} C.~M., {Shields} G.~A., {Wampler} E.~J., 1981, \apj, 249, 443

\bibitem[{{Groves}, {Dopita} \& {Sutherland}(2004{\natexlab{a}}){Groves},
  {Dopita}, \& {Sutherland}}]{groves2004a}
{Groves} B.~A., {Dopita} M.~A., {Sutherland} R.~S., 2004{\natexlab{a}}, \apjs,
  153, 9

\bibitem[{{Groves}, {Dopita} \& {Sutherland}(2004{\natexlab{b}}){Groves},
  {Dopita}, \& {Sutherland}}]{groves2004b}
---, 2004{\natexlab{b}}, \apjs, 153, 75

\bibitem[{{G{\"u}ver} \& {{\"O}zel}(2009)}]{guver2009}
{G{\"u}ver} T., {{\"O}zel} F., 2009, \mnras, 400, 2050

\bibitem[{{Hryniewicz}(2011)}]{hrynio2011}
{Hryniewicz} K., 2011, in Narrow-Line Seyfert 1 Galaxies and their Place in the
  Universe

\bibitem[{{Hryniewicz} {et~al}\mbox{.}(2014){Hryniewicz}, {Czerny}, {Pych},
  {Udalski}, {Krupa}, {{\'S}wi{\c e}to{\'n}}, \& {Kaluzny}}]{hryniewicz2014}
{Hryniewicz} K., {Czerny} B., {Pych} W., {Udalski} A., {Krupa} M., {{\'S}wi{\c
  e}to{\'n}} A., {Kaluzny} J., 2014, \aap, 562, A34

\bibitem[{{Hu} {et~al}\mbox{.}(2008{\natexlab{a}}){Hu}, {Wang}, {Ho}, {Chen},
  {Bian}, \& {Xue}}]{hu2008a}
{Hu} C., {Wang} J.-M., {Ho} L.~C., {Chen} Y.-M., {Bian} W.-H., {Xue} S.-J.,
  2008{\natexlab{a}}, \apjl, 683, L115

\bibitem[{{Hu} {et~al}\mbox{.}(2008{\natexlab{b}}){Hu}, {Wang}, {Ho}, {Chen},
  {Zhang}, {Bian}, \& {Xue}}]{hu2008b}
{Hu} C., {Wang} J.-M., {Ho} L.~C., {Chen} Y.-M., {Zhang} H.-T., {Bian} W.-H.,
  {Xue} S.-J., 2008{\natexlab{b}}, \apj, 687, 78

\bibitem[{{Kaastra} {et~al}\mbox{.}(2011){Kaastra}, {Petrucci}, {Cappi},
  {Arav}, {Behar}, {Bianchi}, {Bloom}, {Blustin}, {Branduardi-Raymont},
  {Costantini}, {Dadina}, {Detmers}, {Ebrero}, {Jonker}, {Klein}, {Kriss},
  {Lubi{\'n}ski}, {Malzac}, {Mehdipour}, {Paltani}, {Pinto}, {Ponti}, {Ratti},
  {Smith}, {Steenbrugge}, \& {de Vries}}]{Kaastra2011}
{Kaastra} J.~S. {et~al.}, 2011, \aap, 534, A36

\bibitem[{{Koshida} {et~al}\mbox{.}(2014){Koshida}, {Minezaki}, {Yoshii},
  {Kobayashi}, {Sakata}, {Sugawara}, {Enya}, {Suganuma}, {Tomita}, {Aoki}, \&
  {Peterson}}]{koshida2014}
{Koshida} S. {et~al.}, 2014, \apj, 788, 159

\bibitem[{{Krolik}, {McKee} \& {Tarter}(1981){Krolik}, {McKee}, \&
  {Tarter}}]{krolik1981}
{Krolik} J.~H., {McKee} C.~F., {Tarter} C.~B., 1981, \apj, 249, 422

\bibitem[{{Landt} {et~al}\mbox{.}(2014){Landt}, {Ward}, {Elvis}, \&
  {Karovska}}]{landt2014}
{Landt} H., {Ward} M.~J., {Elvis} M., {Karovska} M., 2014, \mnras, 439, 1051

\bibitem[{{Leighly}(2004)}]{karen2004}
{Leighly} K.~M., 2004, ApJ, 611, 125

\bibitem[{{Li} {et~al}\mbox{.}(2015){Li}, {Zhou}, {Hao}, {Wang}, {Ji}, {Shi},
  {Liu}, {Zhang}, {Liu}, {Pan}, \& {Jiang}}]{li2015}
{Li} Z. {et~al.}, 2015, \apj, 812, 99

\bibitem[{{Mason}, {Puchnarewicz} \& {Jones}(1996){Mason}, {Puchnarewicz}, \&
  {Jones}}]{mason1996}
{Mason} K.~O., {Puchnarewicz} E.~M., {Jones} L.~R., 1996, \mnras, 283, L26

\bibitem[{{Mathews} \& {Ferland}(1987)}]{Mathews1987}
{Mathews} W.~G., {Ferland} G.~J., 1987, \apj, 323, 456

\bibitem[{{Mehdipour} {et~al}\mbox{.}(2015){Mehdipour}, {Kaastra}, {Kriss},
  {Cappi}, {Petrucci}, {Steenbrugge}, {Arav}, {Behar}, {Bianchi}, {Boissay},
  {Branduardi-Raymont}, {Costantini}, {Ebrero}, {Di Gesu}, {Harrison}, {Kaspi},
  {De Marco}, {Matt}, {Paltani}, {Peterson}, {Ponti}, {Pozo Nu{\~n}ez}, {De
  Rosa}, {Ursini}, {de Vries}, {Walton}, \& {Whewell}}]{Mehdipour2015}
{Mehdipour} M. {et~al.}, 2015, \aap, 575, A22

\bibitem[{{Modzelewska} {et~al}\mbox{.}(2014){Modzelewska}, {Czerny},
  {Hryniewicz}, {Bilicki}, {Krupa}, {{\'S}wi{\c e}to{\'n}}, {Pych}, {Udalski},
  {Adhikari}, \& {Petrogalli}}]{modzelewska2014}
{Modzelewska} J. {et~al.}, 2014, \aap, 570, A53

\bibitem[{{Nenkova} {et~al}\mbox{.}(2008){Nenkova}, {Sirocky}, {Nikutta},
  {Ivezi{\'c}}, \& {Elitzur}}]{nenkova2008}
{Nenkova} M., {Sirocky} M.~M., {Nikutta} R., {Ivezi{\'c}} {\v Z}., {Elitzur}
  M., 2008, \apj, 685, 160

\bibitem[{{Netzer}(1990)}]{netzer1990}
{Netzer} H., 1990, in Active Galactic Nuclei, {Blandford} R.~D., {Netzer} H.,
  {Woltjer} L., {Courvoisier} T.~J.-L., {Mayor} M., eds., pp. 57--160

\bibitem[{{Netzer}(2013)}]{netzer2013}
---, 2013, {The Physics and Evolution of Active Galactic Nuclei}

\bibitem[{{Netzer} \& {Laor}(1993)}]{netzer1993}
{Netzer} H., {Laor} A., 1993, \apjl, 404, L51

\bibitem[{{Osterbrock} \& {Ferland}(2006)}]{osterbrock2006}
{Osterbrock} D.~E., {Ferland} G.~J., 2006, {Astrophysics of gaseous nebulae and
  active galactic nuclei}

\bibitem[{{Peterson} \& {Wandel}(1999)}]{peterson1999}
{Peterson} B.~M., {Wandel} A., 1999, \apjl, 521, L95

\bibitem[{{Pier} \& {Voit}(1995)}]{pier1995}
{Pier} E.~A., {Voit} G.~M., 1995, \apj, 450, 628

\bibitem[{{Rees}, {Netzer} \& {Ferland}(1989){Rees}, {Netzer}, \&
  {Ferland}}]{rees1989}
{Rees} M.~J., {Netzer} H., {Ferland} G.~J., 1989, \apj, 347, 640

\bibitem[{{R{\'o}{\.z}a{\'n}ska} {et~al}\mbox{.}(1999){R{\'o}{\.z}a{\'n}ska},
  {Czerny}, {{\.Z}ycki}, \& {Pojma{\'n}ski}}]{rozanska99}
{R{\'o}{\.z}a{\'n}ska} A., {Czerny} B., {{\.Z}ycki} P.~T., {Pojma{\'n}ski} G.,
  1999, \mnras, 305, 481

\bibitem[{{R{\'o}{\.z}a{\'n}ska} {et~al}\mbox{.}(2006){R{\'o}{\.z}a{\'n}ska},
  {Goosmann}, {Dumont}, \& {Czerny}}]{rozanska2006}
{R{\'o}{\.z}a{\'n}ska} A., {Goosmann} R., {Dumont} A.-M., {Czerny} B., 2006,
  \aap, 452, 1

\bibitem[{{R{\'o}{\.z}a{\'n}ska} {et~al}\mbox{.}(2014){R{\'o}{\.z}a{\'n}ska},
  {Niko{\l}ajuk}, {Czerny}, {Dobrzycki}, {Hryniewicz}, {Bechtold}, \&
  {Ebeling}}]{rozanska2014}
{R{\'o}{\.z}a{\'n}ska} A., {Niko{\l}ajuk} M., {Czerny} B., {Dobrzycki} A.,
  {Hryniewicz} K., {Bechtold} J., {Ebeling} H., 2014, \na, 28, 70

\bibitem[{{Seaton} {et~al}\mbox{.}(1994){Seaton}, {Yan}, {Mihalas}, \&
  {Pradhan}}]{seaton94}
{Seaton} M.~J., {Yan} Y., {Mihalas} D., {Pradhan} A.~K., 1994, \mnras, 266, 805

\bibitem[{{Sredzinska} {et~al}\mbox{.}(2016){Sredzinska}, {Czerny},
  {Hryniewicz}, {Krupa}, {Marziani}, {Adhikari}, {Basak}, {You}, \&
  {Bilicki}}]{sredzinska2016}
{Sredzinska} J. {et~al.}, 2016, ArXiv e-prints

\bibitem[{{Stern} {et~al}\mbox{.}(2016){Stern}, {Faucher-Gigu{\`e}re},
  {Zakamska}, \& {Hennawi}}]{stern2016}
{Stern} J., {Faucher-Gigu{\`e}re} C.-A., {Zakamska} N.~L., {Hennawi} J.~F.,
  2016, \apj, 819, 130

\bibitem[{{Stern}, {Laor} \& {Baskin}(2014){Stern}, {Laor}, \&
  {Baskin}}]{stern2014}
{Stern} J., {Laor} A., {Baskin} A., 2014, \mnras, 438, 901

\bibitem[{{Suganuma} {et~al}\mbox{.}(2006){Suganuma}, {Yoshii}, {Kobayashi},
  {Minezaki}, {Enya}, {Tomita}, {Aoki}, {Koshida}, \&
  {Peterson}}]{suganuma2006}
{Suganuma} M. {et~al.}, 2006, \apj, 639, 46

\bibitem[{{Tanaka} {et~al}\mbox{.}(2014){Tanaka}, {Morokuma}, {Itoh},
  {Akitaya}, {Tominaga}, {Saito}, {Stawarz}, {Tanaka}, {Gandhi}, {Ali}, {Aoki},
  {Contreras}, {Doi}, {Essam}, {Hamed}, {Hsiao}, {Iwata}, {Kawabata}, {Kawai},
  {Kikuchi}, {Kobayashi}, {Kuroda}, {Maehara}, {Matsumoto}, {Mazzali},
  {Minezaki}, {Mito}, {Miyata}, {Miyazaki}, {Mori}, {Moritani},
  {Morokuma-Matsui}, {Morrell}, {Nagao}, {Nakada}, {Nakata}, {Noma}, {Ohsuga},
  {Okada}, {Phillips}, {Pian}, {Richmond}, {Sahu}, {Sako}, {Sarugaku},
  {Shibata}, {Soyano}, {Stritzinger}, {Tachibana}, {Taddia}, {Takaki}, {Takey},
  {Tarusawa}, {Ui}, {Ukita}, {Urata}, {Walker}, \& {Yoshii}}]{tanaka2014}
{Tanaka} M. {et~al.}, 2014, \apjl, 793, L26

\bibitem[{{Thompson}, {Quataert} \& {Murray}(2005){Thompson}, {Quataert}, \&
  {Murray}}]{thompson2005}
{Thompson} T.~A., {Quataert} E., {Murray} N., 2005, \apj, 630, 167

\bibitem[{{van Hoof} {et~al}\mbox{.}(2004){van Hoof}, {Weingartner}, {Martin},
  {Volk}, \& {Ferland}}]{vanhoof2004}
{van Hoof} P.~A.~M., {Weingartner} J.~C., {Martin} P.~G., {Volk} K., {Ferland}
  G.~J., 2004, \mnras, 350, 1330

\bibitem[{{Wills} {et~al}\mbox{.}(1993){Wills}, {Netzer}, {Brotherton}, {Han},
  {Wills}, {Baldwin}, {Ferland}, \& {Browne}}]{wills1993}
{Wills} B.~J., {Netzer} H., {Brotherton} M.~S., {Han} M., {Wills} D., {Baldwin}
  J.~A., {Ferland} G.~J., {Browne} I.~W.~A., 1993, \apj, 410, 534

\bibitem[{{Zhu}, {Zhang} \& {Tang}(2009){Zhu}, {Zhang}, \& {Tang}}]{zhu2009}
{Zhu} L., {Zhang} S.~N., {Tang} S., 2009, \apj, 700, 1173

\end{thebibliography}
\end{document}